\documentclass[a4paper,reqno]{amsart}
\usepackage{amsmath,amssymb,amsfonts,amsthm}
 
\newtheorem{Lemma}{Lemma}
\newtheorem{Theorem}{THEOREM}
\newtheorem{proposition}{Proposition}

\theoremstyle{definition}
\newtheorem{definition}{DEFINITION}
\theoremstyle{definition}
\newtheorem{Remark}{Remark}
\theoremstyle{definition}

\numberwithin{equation}{section}

\newcommand{\R}{\mathbb{R}}

\newcommand{\N}{\mathbb{N}}
\newcommand{\Z}{\mathbb{Z}}
\newcommand{\C}{\mathbb{C}}

\newcommand{\F}{\mathcal{F}}

\newcommand\Pp{\mathcal{P}}
\newcommand{\Hr}{\mathbb{H}}
\newcommand{\gm}{\gamma}
\newcommand{\al}{\alpha}
\newcommand{\ta}{\tilde{\alpha}}
\newcommand{\tg}{\tilde{\gamma}}

\newcommand{\half}{\mbox{$\frac 12$}}
\newcommand{\bra}{\langle}
\newcommand{\ket}{\rangle}

\newcommand{\be}{\begin{equation}}
\newcommand{\ee}{\end{equation}}
\newcommand{\bea}{\begin{align}}
\newcommand{\eea}{\end{align}}

\newcommand\de{\mathcal D}
\newcommand\sgn{{\rm sgn}}

\begin{document}
\title[BCS Functional for General Pair Interactions]
{The BCS Functional for General Pair Interactions}

\author[C. Hainzl]{Christian Hainzl} \address{C. Hainzl, Departments
  of Mathematics and Physics, UAB, 1300 University Blvd, Birmingham AL
  35294, USA} \email{hainzl@math.uab.edu}

\author[E. Hamza]{Eman Hamza}
 \address{E. Hamza, Department of Mathematics, UAB,
   1300 University Blvd, Birmingham AL 35294, USA} \email{hamza@math.uab.edu}

\author[R. Seiringer]{Robert Seiringer}
 \address{R. Seiringer, Department of Physics, Princeton University,
  Princeton NJ 08542-0708, USA} \email{rseiring@math.princeton.edu}

\author[J.P. Solovej]{Jan Philip Solovej}
 \address{J.P. Solovej, Department of Mathematics, University of Copenhagen,
  Universitets\-parken 5, 2100 Copenhagen, Denmark} \email{solovej@math.ku.dk}

\begin{abstract}
  The Bardeen-Cooper-Schrieffer (BCS) functional has recently received
  renewed attention as a description of fermionic gases interacting
  with local pairwise interactions.  We present here a rigorous
  analysis of the BCS functional for general pair interaction
  potentials. For both zero and positive temperature, we show that
  the existence of a non-trivial solution of the {\it nonlinear} BCS
  gap equation is equivalent to the existence of a negative eigenvalue
  of a certain {\it linear} operator. From this we conclude the
  existence of a critical temperature below which the BCS pairing wave
  function does not vanish identically. For attractive potentials, we
  prove that the critical temperature is non-zero and exponentially
  small in the strength of the potential.
\end{abstract}

\date{November 20, 2007}

\maketitle
\renewcommand{\thefootnote}{${}$}
\footnotetext{\copyright\, 2007  by the authors. This paper may be reproduced, in its
entirety, for non-commercial purposes.}

\section{Introduction}

Over the last few years, significant attention has been devoted to
ultra-cold gases of fermionic atoms, both from the experimental
and the theoretical point of view. Because of the ability to
magnetically tune Feshbach resonances it has become possible to
vary interatomic interactions via external magnetic fields, and
hence to control the strength of the interactions among the atoms.
This method was used in remarkable experiments to study the
crossover from the regime of Bose-Einstein condensation (BEC) of
tightly bound diatomic molecules to the Bardeen-Cooper-Schrieffer
(BCS) regime of weakly bound Cooper pairs in ultra-cold gases of
fermionic atoms. For a recent review on this topic we refer to
\cite{zwerger,Chen}.

The main purpose of this paper is to give a mathematically precise
study of the BCS regime. This regime is usually described by the BCS
functional, derived by Leggett \cite{Leg} based on the original work
of Bardeen-Cooper-Schrieffer \cite{BCS}. We note that while in the
original BCS model simple phonon-induced {\em non-local} interaction
potentials were considered, the interaction potentials are {\em local}
for atomic Fermi gases. In order to allow for a wide range of
applications, we will consider here the BCS functional for general pair
interaction potentials $V$.

One purpose of our paper is to obtain necessary and sufficient
conditions on $V$ for the system to display superfluid behavior.
Recall that in the BCS model superfluidity (or, rather,
superconductivity in the original BCS case, where electrons in a
crystal structure have been considered) is related to the
non-vanishing of the pairing wave function, describing particle pairs
with opposite spin and zero total momentum. That is, superfluidity
occurs if the paired state is energetically favored over the normal
state described by the Fermi-Dirac distribution.

More precisely, for arbitrary temperature $T=1/\beta \geq 0$ 
a system is in a superfluid state if there exists a non-trivial solution $\Delta$ to the BCS gap equation
\begin{equation}\label{bcse}
\Delta(p) = -\frac 1{(2\pi)^{3/2}} \int_{\R^3} \hat V(p-q)
\frac{\Delta(q)}{E(q)} \tanh \frac{E(q)}{2T} \, dq\,,
\end{equation}
with $E(p)= \sqrt{(p^2-\mu)^2 + |\Delta(p)|^2}$. Here, $\mu$ denotes the chemical potential.
Notice that the gap equation is highly {\it
  nonlinear}. In Theorem~\ref{thm:minimizer} below we will show that the existence of
a non-trivial solution to the gap equation is equivalent to the
existence of a negative eigenvalue of the corresponding {\it linear}
operator
\begin{equation}\label{tempoperator}
K_{\beta,\mu} +  V\quad ,\quad K_{\beta,\mu} = (p^2 -
\mu)\frac{e^{\beta(p^2-\mu)}+1}{e^{\beta(p^2-\mu)}-1}\,.
\end{equation}
 The operator
$K_{\beta,\mu}$ is understood as a multiplication operator in
momentum space, i.e., $p^2$ equals $-\Delta$ in configuration
space. Observe that in the limit $T\to 0$ this operator reduces to
the Schr\"odinger type operator
\begin{equation}\label{schrtypeoperator}
|p^2 - \mu| + V\,.
\end{equation}
This {\em linear} characterization of the existence of solutions to
the nonlinear BCS equation represents a considerable simplification of
the analysis. In particular, it proves the existence of solutions to
(\ref{bcse}) for a wide range of interaction potentials $V$, and hence
generalizes previous results \cite{billard,bryce,van,yang} valid only for non-local $V$'s under
suitable assumptions.

Moreover, the monotonicity of $K_{\beta,\mu}$ in $\beta$ guarantees
the existence of a {\it critical temperature} $T_c$, with $0\leq T_c <
\infty$, such that superfluidity occurs whenever $T < T_c$ and normal
behavior for $T \geq T_c$. In particular, there is a phase transition
at $T_c$.  For positive values of $\mu$, we shall show that $T_c$ is
strictly positive whenever the negative part of $V$, denoted by $V_-$,
is non-vanishing and the positive part $V_+$ is small enough in a
suitable sense. Moreover, we prove that $T_c$ is exponentially small
in the strength of the interaction potential.  More precisely, if the
potential is given by $\lambda V$ for some fixed $V\in L^1 \cap
L^{3/2}$, then $T_c \lesssim e^{-1/\lambda}$ as $\lambda\to 0$. This
generalizes the corresponding result in \cite[Eq.~(3.29)]{BCS} to a
very large class of interaction potentials. For negative potentials,
also a corresponding lower bound will be shown to hold. In a
forthcoming work \cite{fhss}, the precise asymptotics of $T_c$ for $\lambda\to 0$
will be investigated.

At zero temperature, the existence of a non-trivial solution of
the BCS gap equation is also related to the non-vanishing of the
energy gap between the ground state and the first excited state in
the effective quadratic BCS Hamiltonian on Fock space. For very
specific attractive interaction potentials, it has been argued
that this gap is non-vanishing, see e.g. \cite{Leg,NS,PMTBL}. We
note, however, that the existence of such an energy gap is not
necessarily implied by the non-vanishing of the solution to the
BCS equation. In fact, the presence of a gap is not necessary for
our analysis. We consider general potentials where such an energy
gap need not be present, a priori. We do prove that the gap is
non-vanishing for a certain class of interaction potentials,
however.

We note that our results are relevant for actual experiments on cold
atomic gases as long as the description via the BCS functional is
applicable. There is an extensive literature on this subject, see e.g.
\cite{andre,zwerger,CCPS,Chen,NS,PMTBL,rand,ties}.
According to Leggett \cite{Leg} this is the case for weak interactions
as long as the range of the potential $V$ is much smaller than
$\mu^{-1/2}$ (in appropriate units), i.e., for weakly coupled dilute
gases. In particular, \lq\lq weakly coupled\rq\rq\ means that the
corresponding Schr\"odinger operator $p^2 + V$ typically does {\it
  not} allow for a negative energy bound state since otherwise the
system is in the BEC regime of tightly bound bosonic molecules. We
refer to \cite{Leg} for a detailed discussion on this question.

\section{Model and Main Results}

The precise definition of the BCS functional considered in this paper
is given as follows. For the convenience of reader, we describe in the
appendix the motivation and physical background; the discussion there
follows Leggett's derivation in \cite{Leg}.
See also \cite{BLS} for a detailed study
of the BCS approximation to the Hubbard model.  We use the standard
convention $\hat \alpha(p)=(2\pi)^{-3/2}\int_{\R^3}
\alpha(x)e^{-ipx}dx$ for the Fourier transform.

\begin{definition}\label{defF}
Let $\de$ denote the set of pairs of functions $(\gamma,\alpha)$,
with $\gamma\in L^1(\R^3,(1+p^2)dp)$, $0\leq \gamma(p)\leq 1$, and
$\alpha\in H^1(\R^3,dx)$, satisfying $|\hat\alpha(p)|^2\leq
\gamma(p)(1-\gamma(p))$. Let $V\in L^{3/2}(\R^3,dx)$ be
real-valued and $\mu\in\R$. For $T=1/\beta \geq 0$ and
$(\gamma,\alpha)\in\de$, the energy functional $\F_\beta$ is defined as
\begin{equation}\label{freeenergy}
\F_\beta(\gm,\al)= \int (p^2-\mu)\gm(p)dp+\int |\alpha(x)|^2
V(x)dx -\frac 1\beta S(\gm,\al),
\end{equation}
where $$S(\gm,\al)=-\int \left[ s(p)\ln
  s(p)+\big(1-s(p)\big)\ln\big(1-s(p)\big)\right] dp,$$
with $s(p)$ determined by $s(1-s)=\gm(1-\gm)-|\hat\al|^2$.
\end{definition}

As explained in the Appendix, (\ref{freeenergy}) has the physical
interpretation of $-\half (2\pi)^{3/2}$ times the pressure of a
system of spin $1/2$ fermions at temperature $T$ interacting via a pair potential
given by $2 V$.

It is natural to require $V\in L^{3/2}(\R^3)$, as we do here, since
this guarantees that $V$ is relatively form-bounded with respect to
$p^2$ (see, e.g., Sect.~11 in \cite{LL}). In fact, it would be enough
to require the negative part of $V$ to be in $L^{3/2} +
L^\infty_\epsilon$, and allow for more general positive parts of
$V$. For simplicity of the presentation, we will not pursue this
generalization here.

In the absence of a potential $V$, it is easy to see that
$\F_\beta$ is minimized on $\de$ by the choice $\al\equiv 0$ and
$\gm(p)=\gm_0(p)=[e^{\beta(p^2-\mu)}+1]^{-1}$. We refer to this
state as the {\it normal state} of the system. We will be
concerned with the question of whether a non-vanishing $\alpha$
can lower the energy for given potential $V$.

Note that at $T=0$ it is natural to formulate the model in terms
of a functional that depends only on $\alpha$ and not on $\gamma$.
Namely, for fixed $|\hat\alpha(p)|^2$, the optimal choice of
$\gamma(p)$ is
\begin{equation}\label{gal}
\gamma(p) = \left\{ \begin{array}{ll}
\half (1+\sqrt{1-4|\hat\alpha(p)|^2}) & {\rm for\ } p^2< \mu \\
\half (1-\sqrt{1-4|\hat\alpha(p)|^2}) & {\rm for\ } p^2>\mu
\end{array}\right.
\end{equation}
in this case. Similarly, also at $T>0$ it is enough to minimize
$\F_\beta$ only over pairs $(\gamma,\alpha)$ satisfying a certain
relation independent of $V$, which is displayed in Eq.~(\ref{gap:gamma}) below.

According to the usual interpretation of the BCS theory, the normal
state is unstable if the energy can be lowered by the formation of
Cooper pairs, i.e., if $\F_\beta$ can be lowered by choosing $\alpha
\neq 0$. The following Theorem, which is the main result of this
paper, shows that this property is equivalent to the existence of a
negative eigenvalue of a certain linear operator.

\begin{Theorem}\label{thm:minimizer}
  Let $V\in L^{3/2}(\R^3)$, $\mu\in\R$, and $\infty > T= \frac 1\beta \geq 0$.
  Then the following statements are equivalent:
\begin{itemize}
\item[(i)]
The normal state $(\gamma_0,0)$ is unstable under pair formation,
i.e.,
\begin{equation*}
\inf_{(\gamma,\alpha)\in\de} \F_\beta (\gm,\al) <
\F_\beta(\gm_0,0)\,.
\end{equation*}

\item[(ii)] There exists a pair $(\gamma,\alpha)\in \de$, with $\alpha
  \neq 0$, such that 
\begin{equation}\label{defed}
\Delta(p)=-\dfrac{p^2-\mu}{\gamma(p)-\half}\hat\alpha(p)
\end{equation}
satisfies the BCS gap equation
\begin{equation}\label{bcsgapequationfiniteT}
\phantom{\int\int}
 \Delta=-\hat V \ast \left( \frac{\Delta}{E}\tanh{\dfrac{\beta E}{2}}\right)\ , \ 
{\rm with\ } E(p)=\sqrt{(p^2-\mu)^2+|\Delta(p)|^2}\,.
\end{equation}
\item[(iii)] The linear operator
\begin{equation}\label{linopfinT}
K_{\beta,\mu} +  V\quad ,\quad K_{\beta,\mu} = (p^2 -
\mu)\frac{e^{\beta(p^2-\mu)}+1}{e^{\beta(p^2-\mu)}-1}\,,
\end{equation}
 has at least one negative eigenvalue.
\end{itemize}
\end{Theorem}

Here, $\ast$ denotes convolution in the BCS gap equation
(\ref{bcsgapequationfiniteT}). In order to avoid having to write
factors of $2\pi$, we find it convenient to {\it define}
convolution with a factor $(2\pi)^{-3/2}$ in front, i.e., $(f\ast
g)(p) := (2\pi)^{-3/2} \int_{\R^3}f(p-q)g(q)dq$.

\begin{Remark}
  The operator \eqref{linopfinT} is obtained by taking the second
  derivative of the functional $\F_\beta$ with respect to $\alpha$ at
  the normal state $(\gamma_0,0)$ (see Eq.~(\ref{seca}) below). Since the
  normal state is a critical point of $\F_\beta$, it is then 
  easy to prove that (iii) implies (i). The opposite direction
  is not immediate, however. It says that if $(\gamma_0,0)$ is not a
  minimizer, it can also not be locally stable. 
\end{Remark}

\begin{Remark}
  As mentioned above, a minimizer of $\F_\beta$ satisfies the
  $V$-independent relation (2.2) at $T=0$, and (\ref{gap:gamma}) at
  $T>0$. It is not necessary to assume that these equations hold in
  part (ii) of Theorem~\ref{thm:minimizer}, however.
\end{Remark}

Note that in the zero temperature case $T=0$ the gap equation
(\ref{bcsgapequationfiniteT}) takes the form
$$
\Delta=-\hat V \ast \frac{\Delta}{E}\,,$$
(compare with
\cite{BCS,FW,Leg,MR}) and the operator in \eqref{linopfinT} is given
by $|p^2 - \mu| + V$.

For any potential $V$ we shall prove the existence of a critical
temperature $0\leq T_c = \frac 1\beta_c < \infty$ at which a phase
transition from a normal to superfluid state takes place. More
precisely, we shall show that the gap equation
\eqref{bcsgapequationfiniteT} has a non-trivial solution for all
$T < T_c$, whereas it has no non-trivial solution if $T\geq T_c$.
This follows from the next Theorem, in combination with Theorem
\ref{thm:minimizer}.

\begin{Theorem}\label{phasetransition}
For any $V \in  L^{3/2}(\R^3)$ there exists a critical temperature
$0\leq T_c = \frac 1\beta_c < \infty$ such that
\begin{align*}
\inf {\rm spec} (K_{\beta,\mu} +  V) < 0 \quad & \quad
\mbox{if}\quad T<
T_c\\
K_{\beta,\mu} + V \geq 0 \quad & \quad  \mbox{if} \quad T\geq T_c.
\end{align*}
\end{Theorem}

With the aid of the Birman-Schwinger principle, it is possible to
characterize the value of the critical temperature $T_c$, at least
if $T_c>0$. In order to do this, it is convenient to decompose $V$
into its positive and negative parts, i.e., $V = V_+ - V_-$, with
$V_+\geq 0$, $V_- \geq 0$ and $V_- V_+=0$. We shall show in
Proposition~\ref{criticalbeta} that $T_c>0$ satisfies the equation
\begin{equation}\label{equforcrittemp}
1 = \left\|V_-^{1/2}
  \frac 1{K_{\beta_c,\mu} + V_+}V_-^{1/2}\right \|\,.
\end{equation}
Note that since $K_{\beta,\mu}\geq 2/\beta$, this leads
immediately to the general rough upper bound
\begin{equation}\label{boundoncritical}
 T_c \leq \half\|V_-\|_\infty.
\end{equation}

For negative potentials $V$ and a positive chemical potential
$\mu$, we shall show that the critical temperature is always
positive, i.e., that $|p^2 - \mu| + V$ has a negative energy bound
state. We can even say more, namely we can guarantee superfluidity
for small temperature whenever the positive part $V_+$ is not too
large in a suitable sense.

\begin{Theorem}\label{potandcritical}
Let $V\in L^{3/2}(\R^3)$ be not identically zero, and let $\mu>0$.

\begin{itemize}

\item[(i)]
If $V\leq 0$, then the corresponding critical temperature is
positive, $T_c=\frac 1{\beta_c} > 0$.

\item[(ii)] Let $V = V_+ - V_-$, and let $\beta > \beta_c(-V_-)$,
where $\beta_c(-V_-)$ is the critical inverse temperature for the
potential $-V_-$. Then the system at temperature $1/\beta$ is
still in a superfluid state if
$\|V_+^{1/2}K_{\beta,\mu}^{-1}V_+^{1/2}\|$ is small enough.
\end{itemize}

\end{Theorem}

In particular, this theorem states that for any negative potential
one can add a sufficiently small positive part such that the
critical temperature $T_c$ is still strictly positive.

The next question concerns the magnitude of $T_c$. The rough upper
bound given in (\ref{boundoncritical}) turns out to be much too
rough, in general. In fact, the following Theorem shows that $T_c$
is always exponentially small in the strength of $V$. To be able
to state the bound precisely, let $a:= \inf_{t >0}
t(e^t+1)/((t+2)(e^t-1))$. Numerically, $a\approx 0.654$. Let
further
\begin{equation}\label{deffff}
f(t) := \frac 1{2\pi^2} \int_0^\infty dp\, p^2 \left(
\frac{1}{|p^2-1|+t}-\frac
  1{p^2+1+t}\right)\,.
\end{equation}
Note that $f$ is a positive and monotone decreasing function, with
$f(t) \sim \ln(1/t)$ as $t\to 0$.  We obtain the following upper bound
on the critical temperature $T_c$.

\begin{Theorem}\label{criticaltemperature} Let  $f^{-1}$ be  the
  inverse of the function $f$ in (\ref{deffff}). Let $\mu>0$, and let $V=V_+-V_-$,
  with $V \in L^{3/2}$,
  $V_-\in L^1\cap L^{3/2}$, and with  $\frac 13
\left(\frac 2\pi \right)^{4/3}\|V_-\|_{3/2}< a\approx 0.654$.
Then the critical
  temperature satisfies the upper bound
\begin{equation}
T_c \leq \frac{ \mu  }2 f^{-1} \left( \frac{ a -
    \frac 13
\left(\frac 2\pi
\right)^{4/3}\|V_-\|_{3/2}}{\mu^{1/2}\|V_-\|_1}\right)\,.
\end{equation}
\end{Theorem}
Note that, in particular, $f^{-1}(t) \sim e^{-t}$ for large $t$.
Hence the critical temperature is exponentially small in the potential
for $V_-\in L^1\cap L^{3/2}$.

For negative potentials, one can show that the bound given in
Theorem~\ref{criticaltemperature} is indeed optimal. More
precisely, if the potential is given by $\lambda V$ for some fixed
negative $V \in L^{3/2}$, we shall show in
Proposition~\ref{lowerboundoncrittemp} the lower bound
$$
T_c \geq g_{\mu, V}(\lambda)\,,$$ where $g_{\mu,V}$ is some
positive function depending only on $\mu$ and $V$ satisfying
$g_{\mu,V}(t) \sim e^{-1/t}$ for small $t$. Combining both bounds
then implies that for negative potentials $V$ the critical
temperature is exponentially small in the coupling parameter. This
generalizes well-known calculations in the physics literature, see
e.g. \cite[Eq.~(3.29)]{BCS}, to a very large class of interaction
potentials. Note that, in particular, this result implies that it
is not possible to calculate the critical temperature via a
perturbative expansion in $V$.

\bigskip
Finally, we comment on the continuity properties of minimizers of
$\F_\beta$. In fact, for positive temperature $T>0$ the minimizing
$\gamma$ and $\hat\alpha$ can be shown to be continuous functions.
(We show this in Prop.~\ref{contprop} below.) This may not be
surprising, since already the minimizer without potential,
$\gamma_0(p) = 1/(e^{\beta(p^2 - \mu)} +1)$, is continuous. But in
the case of $T=0$ the situation is not so clear, since the normal
state is a step function with jump at $p^2 = \mu$ for $\mu >0$.

The continuity of a minimizer $\gamma$ at $T=0$ is closely related
to the existence of an {\it energy gap}, given by
\begin{equation}
\Xi := \inf_{p} E(p)\,,
\end{equation}
with $E$ defined in (\ref{bcsgapequationfiniteT}). As explained in the Appendix,
$E(p)$ has the interpretation of the dispersion relation for
quasi-particle excitations. The following shows that in case $V$
has a strictly negative Fourier transform, $\gamma$ is continuous
even at $T=0$ and the gap $\Xi$ is non-vanishing.

\begin{proposition}\label{positivegap}
  Assume that $\inf{\rm spec} (|p^2 - \mu| + V) < 0$, and that $\hat
  V$ is strictly negative. Let $(\gamma,\alpha)\in \de$ be a minimizer
  of $\F_\infty$. Then $\Xi > 0$ and $\gamma$ is a continuous
  function.
\end{proposition}

This result implies, in particular, that $\Xi>0$ for negative $V$
with strictly negative Fourier transform. We note that this property need not necessarily 
be true for all potentials $V$, however.

We conclude this section with two remarks.

\begin{Remark}
  Our results can be generalized to non-local potentials given by an
  integral kernel $V(x,y)$. Exactly such a non-local potential was
  used in the original paper of BCS \cite{BCS} who considered $V$ of
  the form $V(x,y)=-V_0 \overline{\phi(x)}\phi(y)$, with $\hat\phi(p)
  = 1$ if $|p^2 -\mu| \leq h$ and $0$ otherwise. This models an
  effective potential arising from electron-phonon interactions in
  crystals. In this case, the minimizer of $\F_\beta$ can be evaluated
  explicitly (see, e.g., \cite{MR}).

  In particular, our analysis can be used to generalize previous
  results  \cite{billard,bryce,van,yang} on the existence of solutions of the gap equation, as
  already mentioned in the Introduction. 
None of this previous work allows for an interaction
kernel $\hat V(k,k')$ of the form $\hat V(k-k')$, however, since such a kernel does not satisfy appropriate $L^p$ conditions.
Such conditions were necessary for utilization
of fixed point arguments in the previous investigations of this problem.
\end{Remark}

\begin{Remark}
In \cite{CCPS,PMTBL} the BCS-BEC crossover is studied numerically
for certain potentials $V$ that are negative and have negative
Fourier transform. These potentials satisfy all the assumptions of
our results stated above.
\end{Remark}

In the following Sections~\ref{propmin} and ~\ref{tcsect}, the
proofs of the claims of this section will be given.


\section{Properties of the Minimizer of $\F_\beta$}\label{propmin}

In this section we shall give the proof of Theorem
\ref{thm:minimizer} and Proposition~\ref{positivegap} stated
above.  We start by showing the existence of a minimizer of
$\F_\beta$.

\begin{proposition}
There exists a  minimizer of $\F_\beta$ in $\de$.
\end{proposition}

\begin{proof}
We first show that $\F_\beta$ dominates both the
$L^1(\R^3,(1+p^2)dp)$ norm of $\gamma$ and the $H^1(\R^3,dx)$ norm
of $\alpha$. Hence any minimizing sequence will be bounded in
these norms. We have
$$
\F_\beta(\gm,\al) \geq C_1+\frac 34\int (p^2-\mu)\gm(p)dp+\int
|\al(x)|^2 V(x) dx\,,
$$
where
$$ C_1=\inf_{(\gm,\alpha)\in\de}\left( \frac 14\int
(p^2-\mu)\gm(p)dp-\frac 1\beta S(\gm,\alpha) \right)=-\frac1\beta
\int\ln(1+e^{-\frac\beta4(p^2-\mu)})dp\,.
$$
Since $V\in L^{3/2}$ by assumption, it is relatively bounded with
respect to $-\Delta$ (in the sense of quadratic forms), and hence
$C_2=\inf {\rm spec} \, (p^2/4+V)$ is finite. Using $|\hat\alpha(p)|^2\leq \gamma(p)$, we thus have that
$$
\frac 14 \int p^2\gamma(p) dp + \int V(x) |\alpha(x)|^2 dx \geq C_2 \int \gamma(p)dp\,.
$$
Using again that  $|\hat\al(p)|^2 \leq \gamma(p)\leq 1$, it follows
that
\begin{equation}\label{bA}
\F_\beta(\gm,\al)\geq -A+\frac18
\|\al\|_{H^1(\R^3,dx)}^2+\frac{1}{8}\|\gm\|_{L^1(\R^3,(1+p^2)dp)}\,,
\end{equation}
where
$$
A=-C_1- \int \left[p^2/4 - 3 \mu/4 - 1/4 + C_2\right]_-dp\,,
$$
with $[\,\cdot\,]_- = \min\{\,\cdot\,,0\}$ denoting the negative part.

To show that a minimizer of $\F_\beta$ exists in $\de$, we pick a
minimizing sequence $(\gm_n,\al_n)\in\de$, with
$\F_\beta(\gm_n,\al_n)\leq 0$. From (\ref{bA}) we conclude that
$\|\al_n\|_{H^1}^2\leq 8A$, and hence we can find a subsequence
that converges weakly to some $\ta\in H^1$. Since $V\in
L^{3/2}(\R^3)$, this implies that
$$
\lim_{n\to\infty} \int_{\R^3} V(x)|\al_n(x)|^2 dx = \int_{\R^3}
V(x)|\tilde\al(x)|^2 dx
$$
\cite[Thm.~11.4]{LL}.

It remains to show that the remaining part of the functional,
\begin{equation}\label{f0}
\F^0_\beta(\gm,\al)=\int \left[ (p^2-\mu)\gm(p)dp + \frac 1\beta
  s(p)\ln s(p) + \frac 1 \beta \big(1-s(p)\big) \ln\big(1-s(p)\big)
\right]dp\,,
\end{equation}
is weakly lower semicontinuous. Note that $\F^0_\beta$ is {\it
jointly
  convex} in $(\gamma,\alpha)$, and that its domain $\de$ is a convex
set.  We already know that $\alpha_n\rightharpoonup \ta$ weakly in
$H^1(\R^3)$. Moreover, since $\gm_n$ is uniformly bounded in
$L^1(\R^3)\cap L^\infty(\R^3)$, we can find a subsequence such
that $\gm_n \rightharpoonup \tg$ weakly in $L^p(\R^3)$ for some
$1<p<\infty$. We can then apply Mazur's theorem
\cite[Theorem~2.13]{LL} to construct a new sequence as convex
combinations of the old one, which now converges {\it strongly} to
$(\tg,\ta)$ in $L^p(\R^3)\times L^2(\R^3)$.  By going to a
subsequence, we can also assume that $\gamma_n\to \gamma$ and $\hat
\alpha_n \to \hat {\tilde \alpha}$ pointwise
\cite[Theorem~2.7]{LL}. Because of convexity of $\F^0_\beta$, this
new sequence is again a minimizing sequence.

Note that the integrand in (\ref{f0}) is bounded from below
independently of $\gamma$ and $\alpha$ by $-\beta^{-1}
\ln(1+e^{-\beta(p^2-\mu)})$. Since this function is integrable, we
can apply Fatou's Lemma \cite[Lemma~1.7]{LL}, together with the
pointwise convergence, to conclude that $\liminf
\F^0_\beta(\gm_n,\al_n) \geq \F^0_\beta(\tg,\ta)$.

We have thus shown that
$$
\F_\beta(\tg,\ta)\leq \liminf_{n\to\infty}
\F_\beta(\gm_n,\al_n)\,.
$$
It is easy to see that $(\tg,\ta)\in\de$, hence it is a minimizer.
This proves the claim.
\end{proof}


Next, we establish a few auxiliary results that will be needed
below in the proof of Theorem \ref{thm:minimizer}.

\begin{Lemma}\label{l:gapequation}
  Let $(\gm,\al)\in \de$ be a minimizer of $\F_\beta$ for $0<\beta<\infty$. Then, for a.e. $p\in\R^3$,
 \begin{align}
   (\hat V\ast \hat\al)(p) &=(p^2-\mu)\frac{\hat\al(p)}{2\gm(p)-1}
   \,,\label{gap:alpha} \\ (p^2-\mu)&=-\frac {2\gm(p)-1}{\beta}
   f\left(2\sqrt{(\gm(p)-\half)^2+|\hat\al(p)|^2}\right)\label{gap:gamma}\,,
\end{align}
where $f(a):=\frac 1a \ln{\frac{1+a}{1-a}}$ for $0\leq a<1$.
\end{Lemma}

\begin{proof}
  For $\epsilon>0$, let
  $A_\epsilon:=\{p\in\R^3:|\hat\al(p)|^2+(\gm(p)-1/2)^2
  \leq 1/4-\epsilon\}$. Let $g\in H^1(\R^3)\cap L^1(\R^3)$,
  with Fourier transform supported in $A_\epsilon$. Then
  $(\gamma,\alpha+t g)\in \de$ for small enough $|t|$. By the Lebesgue
  dominated convergence theorem, it follows that
\begin{align}\nonumber
\left. \frac{d}{dt} \F_\beta(\gm,\al+tg) \right|_{t=0}&= 2\, {\rm
Re\, }\int V(x)\overline{ g(x)}\alpha(x)\,dx \\ &\quad
+\frac2\beta\, {\rm Re\, } \int \overline {\hat g(p)}
\hat\alpha(p)
f\left(2\sqrt{(\gm(p)-\half)^2+|\hat\al(p)|^2}\right)dp\,.
\label{e2}
\end{align}
Similarly, for $\hat g$ real, $(\gamma+t\hat g,\alpha)\in\de$ for
small $|t|$, and we get
\begin{align}\label{e1}
\left. \frac{d}{dt} \F_\beta(\gm+t\hat g,\al)\right|_{t=0}&=\int
(p^2-\mu)\hat g(p) dp\\ &\quad+\frac1\beta\int \hat
g(p)(2\gm(p)-1)
f\left(2\sqrt{(\gm(p)-\half)^2+|\hat\al(p)|^2}\right)dp \,.
\nonumber
\end{align}
Since $(\gm,\al)$ minimize $\F_\beta$ by assumption, the
expressions in (\ref{e2}) and (\ref{e1}) vanish. It follows that
for a.e. $p\in A_\epsilon$ we have
\begin{align}
 (\hat V\ast
\hat\al)(p)&=-\frac1\beta\hat\al(p)f\Big(2\sqrt{(\gm(p)-\half)^2+|\hat\al(p)|^2}\Big)\,,\nonumber
\\ (p^2-\mu)&= -\frac1\beta(2\gm(p)-1)f\Big(2\sqrt{(\gm(p)-\half)^2+|\hat\al(p)|^2}\Big)\,.\label{gamma}
\end{align}

To conclude that (\ref{gamma}) holds a.e. in $\R^3$, it remains to
show that $B:=\R^3 \setminus \cup_{\epsilon>0} A_\epsilon$ has
zero measure. By definition, if $p\in B$ we have
$\gm(p)(1-\gm(p))=|\hat\al(p)|^2$, i.e. $s(p)\big(s(p)-1\big)=0$.
It is then easy to see that if $B$ has non-zero measure,
$\F_\beta(\gamma,\alpha)$ can be lowered by modifying $\gm$ and
$\hat\al$ on $B$ (since $t\mapsto t\ln t$ has a diverging
derivative at zero). This contradicts the fact the
$(\gamma,\alpha)$ is a minimizer, and hence $B$ has zero measure.
\end{proof}

\begin{Remark}\label{remt0}
  A similar strategy as in the proof of Lemma~\ref{l:gapequation}
  leads to the conclusion that Eq.~(\ref{gap:alpha}) is also valid at
  $T=0$. The relation between $\gamma$ and $\alpha$, given in
  Eq.~(\ref{gap:gamma}) for $T>0$, has to be replaced by (\ref{gal}) at
  $T=0$, however. We omit the details.
\end{Remark}

Note that Eq. (\ref{gap:gamma}) and the positivity of $f$
immediately imply that $\gamma(p)\geq 1/2$ for $p^2<\mu$, and
$\gamma(p)\leq 1/2$ for $p^2>\mu$. Of greater importance below
will be the following monotonicity property, which bounds $\gamma$
in the opposite direction.

\begin{Lemma}\label{monotone}
  Assume that  $(\gm,\al)\in\de$ satisfies Eq.~(\ref{gap:gamma}), and let
  $\gamma_0(p)=[e^{\beta(p^2-\mu)}+1]^{-1}$.  For a.e.  $p\in\R^3$, we
  have $\gm(p)\leq\gm_0(p)$ if $p^2<\mu$, while for $p^2>\mu$ we have
  $\gm(p)\geq\gm_0(p)$.  Moreover, these inequalities are strict if
  $\hat\al(p)\neq 0$.  In particular,
  $\dfrac{p^2-\mu}{1-2\gm_0(p)}<\dfrac{p^2-\mu}{1-2\gm(p)}$ on the
  support of $\hat \alpha$.
\end{Lemma}

\begin{proof}
Let $p^2>\mu$ such that $\hat\al(p)\neq0$ and assume by way of
contradiction that $\gm(p)\leq\gm_0(p)$. We then have
$$\frac 12 -\gm_0(p) < \sqrt{|\hat\al(p)|^2+(\half-\gm(p))^2}\,.$$
Since $f$ is a strictly monotone increasing function, we obtain
$$ f\big(1-2\gm_0(p)\big) < f\left(2\sqrt{|\hat\al(p)|^2+(\half-\gm(p))^2}\right)$$
and hence
\begin{equation}\label{inequalitycontradiction}
(1-2\gm_0(p))f\big(1-2\gm_0(p)\big)<(1-2\gm(p))f\left(2\sqrt{|\hat\al(p)|^2+(\half-\gm(p))^2}\right)
\,.
\end{equation}
This, however, contradicts the fact that both $(\gm_0,0)$ and
$(\gm,\al)$ are solutions of \eqref{gap:gamma}, and hence both
sides of \eqref{inequalitycontradiction} are equal to
$\beta(p^2-\mu)$.

The case $p^2<\mu$ is treated similarly.
\end{proof}


We now have all the necessary prerequisites at our disposal to
give the proof of Theorem~\ref{thm:minimizer}.

\begin{proof}[Proof of Theorem \ref{thm:minimizer}]
  [i] $\Rightarrow$ [ii]: Consider first the case $T>0$. According to
  Lemma~\ref{l:gapequation} a minimizing pair $(\gm,\al)\in\de$ with
  $\al\neq 0$ satisfies Eqs.~(\ref{gap:alpha}) and
  (\ref{gap:gamma}). Using the definitions of $\Delta$ and $E$ in
  \eqref{defed} and (\ref{bcsgapequationfiniteT}) we see that
  $\sqrt{(\gamma(p) - 1/2)^2 + |\hat \alpha(p)|^2} = \frac{1/2-
    \gamma(p)}{p^2 - \mu}E(p)$. Plugging this into
  Eq.~(\ref{gap:gamma}) leads to
$$
\frac{p^2-\mu}{2\gamma(p)-1} = -\frac {1}{\beta} f \left(
  \frac{1-2\gamma(p)}{p^2-\mu}E(p)\right) = \frac 2{\beta E(p)} \frac{
  p^2-\mu}{2\gamma(p)-1} \tanh^{-1}\! \left(
  \frac{1-2\gamma(p)}{p^2-\mu}E(p)\right),
$$
or
\begin{equation}\label{t1e1}
\tanh{\frac{\beta
E(p)}{2}} = \dfrac{1- 2\gamma(p)}{p^2-\mu} E(p)\,.
\end{equation}
Therefore, we can rewrite the relation between $\hat \alpha$ and
$\Delta$ as $\hat \alpha(p) = \frac{\Delta(p)}{2E(p)}
\tanh{\frac{\beta E(p)}{2}}$. Together with Eq.~(\ref{gap:alpha})
this gives the required form of the gap equation.

The case $T=0$ is treated similarly (cf. Remark~\ref{remt0}). 

[ii] $\Rightarrow$ [iii]: Given $T\geq 0$ and a $\Delta$ satisfying
(\ref{defed}) for some $(\gamma,\alpha)\in\de$, we first show that
there exists a $(\tilde\gamma,\tilde\alpha)\in \de$ yielding the same
$\Delta$ but satisfying in addition Eq.~(\ref{gal}) at $T=0$ or
Eq.~(\ref{gap:gamma}) at $T>0$, respectively. Namely, we can let
$\hat{\tilde\alpha}(p) = m(p)\hat\alpha(p)$ and $\tilde\gamma(p)=1/2 +
m(p)(\gamma(p)-1/2)$ without changing $\Delta$. If we choose
$$
m(p) = \sgn\left(\frac{p^2-\mu}{1-2\gamma(p)}\right) 
\frac{\tanh(\beta E(p)/2)}{2\sqrt{(\gamma(p)-\half)^2+|\hat\alpha(p)|^2}}\,,
$$
with $E(p)$ as in (\ref{bcsgapequationfiniteT}), it is easy to see
that (\ref{t1e1}) is satisfied. Using
in addition the gap equation, the same reasoning as in the previous
part of the proof leads to the conclusion that Eqs. (\ref{gap:alpha})
and (\ref{gal}) (respectively (\ref{gap:gamma})) are 
satisfied for this $(\tilde\gamma,\tilde\alpha)$.

Hence there exists a $(\gamma,\alpha)\in \de$ satisfying
(\ref{gap:alpha}) as well as (\ref{gal}) (at $T=0$) or (\ref{gap:gamma}) (at
$T>0$).  Using (\ref{gap:alpha}) and the definition of
$K_{\beta,\mu}$, we have that
\begin{align*}
\bra\al,(K_{\beta,\mu}+V)\al\ket=\left\bra\al,\left(\frac
{1}{1-2\gm_0(p)}-\frac{1}{1-2\gamma(p)}\right)(p^2-\mu)\al\right\ket\,.
\end{align*}
It follows from Lemma \ref{monotone} that this expression is negative
whenever $\alpha$ is not identically zero.

 [iii] $\Rightarrow$ [i]:
Consider first the case $T>0$. Showing that the existence of a
negative eigenvalue of the operator \eqref{linopfinT} implies that
the minimizing $\al$ is not identically zero is equivalent to
proving that $\al\equiv 0$ implies that $\inf{\rm
spec\,}(K_{\beta,\mu}+ V)\geq0$. First, we note that, for any
$\delta\in C^\infty_0$, $\|(t\delta)^2+(\gm_0-\frac
12)^2\|_\infty<\frac 14$ for all small $|t|$. With the aid of the
Lebesgue dominated convergence theorem it is easy to see that
$\frac{d^2}{dt^2}\F_\beta(\gm_0,t\hat\delta)$ exists for small
$t$, and is given by
\begin{align*}
  &\frac{d^2}{dt^2}\F_\beta(\gm_0,t\hat\delta) \\ & = 2 \int V(x)
  |\hat\delta(x)|^2 \, dx\\& \quad +\frac 2\beta\int
  |\delta(p)|^2\frac{(\gm_0(p)-\frac
    12)^2}{|t\delta(p)|^2+(\gm_0(p)-\half)^2} f\left(2
    \sqrt{|t\delta(p)|^2+(\gm_0(p)-\half)^2}\right)dp \\& \quad +\frac
  1\beta\int\frac{t^2|\delta(p)|^4}
  {[|t\delta(p)|^2+(\gm_0(p)-\half)^2][\gm_0(p)(1-\gm_0(p))-|t\delta(p)|^2]}\,dp\,.
\end{align*}
Assuming that the minimizing $\al$ is identically zero, it follows
that the corresponding $\gm=\gm_0$ and hence
$\frac{d^2}{dt^2}\F_\beta(\gm_0,t\hat\delta)|_{t=0}\geq 0$. But
\begin{equation}\label{seca}
\left. \frac{d^2}{dt^2}\F_\beta(\gm_0,t\hat\delta)\right|_{t=0}=
2\int V(x)|\hat\delta(x)|^2 \, dx + 2\int |\delta(p)|^2
K_{\beta,\mu}(p) \, dp \,,
\end{equation}
and hence $\bra \delta, (K_{\beta,\mu}+ V)\delta\ket\geq0$ for all
$\delta\in C_0^\infty$. This proves the statement.

In the case $T=0$, it is sufficient to minimize $\F_\beta$ over
$(\gm,\al)\in\de$ with $\gamma$ of the form (\ref{gal}), as
remarked earlier. For simplicity, we abuse the notation slightly
and denote the resulting functional by $\F_\infty(\alpha)$.
Subtracting its value for $\alpha=0$, it is given by
\begin{equation}
 \F_\infty(\alpha) -\F_\infty(0)
=\frac 12 \int|p^2-\mu|(1-\sqrt{1-4|\hat\al(p)|^2})dp+ \int
V(x)|\alpha(x)|^2\,dx\,.
\end{equation}
Assume that $|p^2-\mu|+V$ has a negative eigenvalue. We can then
find an $\al\in C_0^\infty(\R^3)$ such that $\bra\al,(|p^2-\mu|+
V)\al\ket<0$. For $0<\epsilon\ll 1$ small enough, we then have
\begin{equation}
\F_\infty(\epsilon\al)-\F_\infty(0) =
 \epsilon^2\bra\al,(|p^2-\mu|+ V)\al\ket+\mathcal{O}(\epsilon^4)<0\,.
\end{equation}
This implies that a minimizer $\al$ does not vanish identically.
\end{proof}


In the following, we investigate the continuity properties of a
minimizing pair $(\gamma,\alpha)\in \de$.

\begin{proposition}\label{contprop}
If $(\gamma,\al)\in\de$ is a minimizer of $\F_\beta$ for
$0<\beta<\infty$, then both $\gamma$ and $\hat\al$ are continuous
functions.
\end{proposition}

\begin{proof}
  Since $V\in L^{3/2}$ and $\al\in H^1$, we know that $\hat V \in L^3$
  and $\hat\alpha \in L^{6/5+\epsilon}$ for all $\epsilon>0$. Hence
  $\hat V\ast \hat \alpha$ is continuous \cite[Thm.~2.20]{LL}. From
  Eq.~(\ref{gap:alpha}) and the definition of $\Delta$ and $E$ given
  in (\ref{defed}) and (\ref{bcsgapequationfiniteT}), we observe that
  $\hat V\ast\hat\al=-\half \Delta$ and hence both $\Delta$ and $E$
  are continuous functions. As in Eq.~(\ref{t1e1}) in the proof of
  Thm.~\ref{thm:minimizer}, Eq.~(\ref{gap:gamma}) can be rewritten in
  the form
\begin{equation}\label{E:gap}
\dfrac{p^2-\mu}{2\gamma(p)-1}=-\dfrac{e^{\beta E(p)}+1}{e^{\beta
E(p)}-1}E(p)\,.
\end{equation}
The right side of this equation is continuous and does not vanish, not
even when $E(p)=0$ (which could happen, in principle, for $p^2=\mu$).
Hence $\gamma$ is continuous. Using again Eq.~(\ref{gap:alpha}), this
also implies continuity of $\hat\alpha$.
\end{proof}


The situation is different at zero temperature, however. In the
limit $\beta\to \infty$, the right side of Eq.~(\ref{E:gap})
vanishes in case $E(p)=0$. Hence $\gamma$ could possibly be
discontinuous if $p^2=\mu$. In fact, this is what happens in the
case of vanishing potential. On the other hand, a non-vanishing
gap $\Xi = \inf_p E(p)>0$ implies continuity of $\gamma$.  At
$T=0$, the gap equation (\ref{bcsgapequationfiniteT}) takes the
form
\begin{equation}\label{gapatzero}
E(p)\hat\al(p)=-(\hat{V}\ast\hat\al)(p)\,,
\end{equation}
with
$E(p)=\dfrac{|p^2-\mu|}{\sqrt{1-4|\hat\al(p)|^2}}$. Hence the gap
$\Xi$ is guaranteed to be non-zero if the right side of
(\ref{gapatzero}) vanishes nowhere on the sphere $p^2=\mu$. This
can be easily shown in case $\hat V$ is strictly negative, which
is the content of our Proposition~\ref{positivegap}.

\begin{proof}[Proof of Proposition~\ref{positivegap}]
Since $\hat V\leq0$,
\begin{equation}\label{sta}
\int \overline{ \hat\alpha}(\hat V\ast\hat\alpha)\geq \int
|\hat\alpha|(\hat V\ast|\hat\alpha|)\,.
\end{equation}
Moreover, since $\hat V$ is assumed to be strictly negative, there
can be equality in (\ref{sta}) only if $\hat\alpha(p)=
e^{i\kappa}|\hat\alpha(p)|$ for some constant $\kappa\in\R$. This
implies that this is the case for a minimizer of $\F_\infty$, and
hence $\hat V\ast \hat\alpha$ vanishes nowhere. The statement then
follows from Eq.~(\ref{gapatzero}).
\end{proof}


\section{The Critical Temperature}\label{tcsect}

In this section we investigate the relation between properties of
the interaction potential $V$ and the critical temperature $T_c$
at which the phase transition takes place. In particular, we shall
prove Theorems~\ref{phasetransition}--\ref{criticaltemperature}.

\begin{proof}[Proof of Theorem~\ref{phasetransition}]
Let $\beta_c=\inf_\beta \{\beta>0:\inf {\rm
spec}(K_{\beta,\mu}+V)<0\}$. It is easy to see that $\beta_c>0$, since
$V$ is relatively form-bounded with respect to $|p^2-\mu|$ and
$K_{\beta,\mu}\geq c(|p^2-\mu|+1/\beta)$ for some $c>0$
(cf. Lemma~\ref{l11} below).

From the definition of $K_{\beta,\mu}$ in (\ref{linopfinT}) one easily
sees that 
\begin{equation}
K_{\beta_1,\mu}\leq K_{\beta_2,\mu} \qquad \text{if  $\beta_1\geq
\beta_2$.}
\end{equation}
 It follows that $\inf {\rm spec}(K_{\beta,\mu}+ V)<0$ for all
 $\beta>\beta_c$. Moreover, since $K_{\beta,\mu}$ depends continuously on
 $\beta$ (in a norm resolvent sense), we have that
 $K_{\beta_c,\mu}+ V\geq 0$. Setting $\beta_c=\infty$ in the case
 $\inf{\rm spec}(K_{\beta,\mu}+ V)\geq0$ for all $\beta$ and
 letting $T_c=1/\beta_c$ finishes the proof.
\end{proof}

Next, we prove the first assertion of Theorem
\ref{potandcritical}, namely that for any negative potential the
critical temperature is strictly positive.

\begin{proof}[Proof of Theorem \ref{potandcritical}(i)]
In light of Theorem \ref{phasetransition} it suffices to prove
that $\inf {\rm spec}(|p^2-\mu|+V)<0$. According to the
Birman-Schwinger principle, this is equivalent to the fact that
for some $e>0$, the Birman-Schwinger operator
\[B_{e}:=|V|^{1/2}(|p^2-\mu|+e)^{-1}|V|^{1/2}\] has an eigenvalue $1$.
Notice that $\lim_{e\to\infty} \|B_{e}\| = 0$ since $V\in L^{3/2}$
is relatively bounded with respect to $p^2$ (in the sense of
quadratic forms). Because of continuity in $e$, it thus suffices
to find an $e>0$ such that $\|B_{e}\|>1$.

Hence we have to show that there exists a $\phi\in L^2(\R^3)$ with
$\|\phi\|_2=1$ such that $\bra \phi,B_{e}\phi\ket\geq 1$ for an
appropriate $e$. For this purpose, we choose a normalized function
$\phi$ in such a way that $\widehat{\phi|V|^{1/2}}$ does not vanish in
a neighborhood of some point $p$ with $p^2=\mu$. We can, for
instance, choose $\phi$ to be rapidly decaying, in which case
$|\widehat{\phi|V|^{1/2}}|$ is continuous and hence strictly positive on
some small ball. This ball can then be made to overlap with the sphere
$p^2=\mu$ by simply multiplying $\phi$ by a factor $e^{iqx}$ for
appropriate $q$, which translates the function
$\widehat{\phi|V|^{1/2}}$ by $q$. In this way,
\[ \left\bra \phi, B_e \phi\right\ket =
\int |\widehat{\phi|V|^{1/2}}(p)|^2\dfrac{1}{|p^2-\mu|+e}\, dp \to
\infty \quad {\rm as}\quad  e\to 0.
\]
Thus there exists $e>0$ such that $\bra \phi,B_{e}\phi\ket>1$.
\end{proof}

Before proceeding with the proof of the second part of Theorem
\ref{potandcritical}, we introduce an alternative characterization
for the critical temperature $T_c$ in the case $T_c>0$, which we
already referred to in Eq.~(\ref{equforcrittemp}).

\begin{proposition}\label{criticalbeta}
For all $V \in L^{3/2}(\R^3)$, such that $V=V_+-V_-$, with
$V_+\geq0$, $V_-\geq0$ and $V_-V_+=0$, the critical temperature
$T_c= 1/\beta_c$ is characterized by
\[\left\|V_-^{1/2} \frac 1{K_{\beta_c,\mu} + V_+}V_-^{1/2}\right
\|=1\,,
\]
whenever $T_c\neq 0$.
\end{proposition}

Recall that $K_{\beta,\mu}\geq 2/\beta$, hence $K_{\beta,\mu}+V_+$ has
a bounded inverse.

\begin{proof}
Let the operator $B^\beta_{e}$ be defined as
\[B^\beta_{e}=V_-^{1/2} \frac 1{K_{\beta,\mu} +
  V_++e}V_-^{1/2}.\] This operator is compact (in fact,
Hilbert-Schmidt) and depends continuously on $\beta$ and $e$.
Moreover, $\|B_{e}^\beta\|$ is strictly decreasing in $e$ and
$1/\beta$. 

Since $T_c>0$ by assumption, we know from
Theorem~\ref{phasetransition} that $K_{\beta,\mu} + V$ has a negative
eigenvalue for $T<T_c$. Hence the Birman-Schwinger principle implies
that $\|B_e^\beta\|>1$ for $1/\beta$ and $e$ small enough. On the
other hand, $\|B_0^\beta\| \to 0$ as $\beta \to 0$. Thus there exists
a unique $\tilde\beta$ such that $\|B^{\tilde{\beta}}_{0}\|=1$.

For $\beta > \tilde\beta$, $\|B_e^\beta\|>1$ for $e$ small enough,
and hence $K_{\beta,\mu}+V$ has a negative energy bound state. For
$\beta < \tilde\beta$, however, $\|B_e^\beta\|<1$ for all $e\geq
0$, and hence $K_{\beta,\mu} + V \geq 0$. This shows that
$\tilde\beta = \beta_c$.
\end{proof}

From Prop.~\ref{criticalbeta} it is then easy to conclude that the
critical temperature stays strictly positive under small positive
perturbations to negative potentials.

\begin{proof}[Proof of Theorem \ref{potandcritical}(ii)]
  According to Prop.~\ref{criticalbeta}, it suffices to prove that for
  $\beta>\beta_c(-V_-)$ and for $\lambda>0$ small enough,
  $\|V_-^{1/2}(K_{\beta,\mu}+\lambda V_+)^{-1}V_-^{1/2}\|>1$.
  In order to simplify the notation let
  $B_\lambda=V_-^{1/2}(K_{\beta,\mu}+\lambda V_+)^{-1}V_-^{1/2}$. It follows that
\begin{align}\nonumber
   B_\lambda &= V_-^{1/2} \frac
  1{K_{\beta,\mu}^{1/2}} \left(\frac 1 {1+ \lambda
      K_{\beta,\mu}^{-1/2} V_+ K_{\beta,\mu}^{-1/2}}
  \right) \frac 1{K_{\beta,\mu}^{1/2}} V_-^{1/2} \\ \nonumber
  &\geq  B_0 \, \left(1+ \lambda \| K_{\beta,\mu}^{-1/2} V_+
  K_{\beta,\mu}^{-1/2} \|\right)^{-1}\,.
\end{align}
Since $V_+\in L^{3/2}$ by assumption, it is relatively
form-bounded with respect to $K_{\beta,\mu}$, and hence  $\|
K_{\beta,\mu}^{-1/2} V_+ K_{\beta,\mu}^{-1/2} \|
=\|V_+^{1/2}K_{\beta,\mu}^{-1}V_+^{1/2}\|$ is finite. Therefore
$\|B_\lambda\| > 1$ if $\|B_0\| >1$ and  $\lambda$ is small
enough.
\end{proof}

Next, we shall derive upper and lower bounds on the critical
temperature. These bounds are similar to the ones found in the
literature, but generalize these results from simple rank 1
potentials as in \cite{BCS} to very general pair interactions.

We start with the upper bound given in
Theorem~\ref{criticaltemperature}. To this aim we first prove two
auxiliary lemmas.

\begin{Lemma}\label{l11}
Let $K_{\beta,\mu}(p) = (p^2 -
\mu)(e^{\beta(p^2-\mu)}+1)(e^{\beta(p^2-\mu)}-1)^{-1}$. Then there
is a constant $a>0$ such that
\begin{equation}\label{uplowboundK}
a \left( |p^2-\mu| + \frac 2\beta\right) \leq K_{\beta,\mu}(p)
\leq \left( |p^2-\mu| + \frac 2\beta\right)\,.
\end{equation}
\end{Lemma}

\begin{proof}
  Let $g(t)=t(e^t+1)/((t+2)(e^t-1))$ for $t\geq 0$. It is easy to see
  that $g(t)\leq 1$. Moreover, $g(0)=1$ and $\lim_{t\to\infty}g(t)=1$,
  hence $a:=\inf_{t>0}g(t)>0$. In fact, $a\approx 0.654$ numerically.
  Since $K_{\beta,\mu}(p)=(|p^2-\mu|+2/\beta)g(\beta|p^2-\mu|)$, this
  implies the statement.
\end{proof}

\begin{Lemma}\label{l12} Assume that $V\in L^1\cap L^{3/2}$ and that
  $\mu>0$.
Then
\begin{equation}
\left\| |V|^{1/2}  \frac 1{|p^2-\mu|+e} |V|^{1/2}\right\| \leq
\mu^{1/2}\|V\|_1  f(e/\mu) +\frac 13 \left(\frac 2\pi
\right)^{4/3} \|V\|_{3/2}\,,
\end{equation}
where $f$ is defined in (\ref{deffff}).
\end{Lemma}

\begin{proof}
Define $k_{e,\mu}(p)\geq 0$ by
\begin{equation}\label{kme}
\frac 1{|p^2-\mu|+e} = \frac 1{p^2+\mu+e} + k_{e,\mu}(p)\,.
\end{equation}
Then
\begin{equation}
\left\| |V|^{1/2}  \frac 1{|p^2-\mu|+e} |V|^{1/2}\right\|  \leq
\left\|    |V|^{1/2}  \frac 1{p^2+\mu+e} |V|^{1/2}\right\|  \label{dec}  + \left\|
    |V|^{1/2}
k_{e,\mu}(p) |V|^{1/2}\right\|\,.
\end{equation}
The last term is bounded by
\begin{align*}
\left\| |V|^{1/2} k_{e,\mu}(p) |V|^{1/2}\right\| =
&\sup_{\|\psi\|_2=1} \langle \psi |V|^{1/2},\hat k_{e,\mu} \ast
(\psi |V|^{1/2})\rangle \\ & \leq (2\pi)^{-3/2} \|\hat
k_{e,\mu}\|_\infty \|V\|_1 \leq (2\pi)^{-3} \|k_{e,\mu}\|_1
\|V\|_1\,.
\end{align*}
Moreover, $\|k_{e,\mu}\|_1 = (2\pi)^3\mu^{1/2} f(e/\mu)$, with $f$
given in (\ref{deffff}).

For the first term on the right side of (\ref{dec}), the use of
the Hardy-Littlewood-Sobolev inequality and H\"older's inequality
yields
\begin{equation}\nonumber
\left\|
    |V|^{1/2}  \frac 1{p^2+\mu+e} |V|^{1/2}\right\| \leq \sup_{\|\psi\|_2=1} \left\langle
\psi |V|^{1/2},\frac 1{p^2} \psi |V|^{1/2}\right\rangle \leq \frac
13 \left(\frac 2\pi \right)^{4/3}\|V\|_{3/2}\,.
\end{equation}
\end{proof}

Equipped with these two lemmas we are now able to prove the upper
bound on the critical temperature stated in Theorem
\ref{criticaltemperature}.

\begin{proof}[Proof of Theorem \ref{criticaltemperature}]
Assume that $T_c>0$, for otherwise there is nothing to prove.
According to Proposition~\ref{criticalbeta}, the critical
temperature $T_c=1/\beta_c$ is determined by the equation
\begin{equation}\nonumber
\left\| V_-^{1/2} \frac 1{K_{\beta_c,\mu}+ V_+} V_-^{1/2}\right\|
= 1\,.
\end{equation}
Using Lemmas~\ref{l11} and~\ref{l12} as well as $V_+\geq 0$, we
thus have
\begin{align}\nonumber
1 &\leq \frac 1a \left\| V_-^{1/2} \frac 1{|p^2-\mu|+2/\beta_c}
  V_-^{1/2}\right\| \\ &\leq \frac{\mu^{1/2}\|V_-\|_1}a
f\big(2/(\beta_c\mu)\big) +  \frac 1{3a} \left(\frac 2\pi
\right)^{4/3}\|V_-\|_{3/2}\,. \nonumber
\end{align}
Since $f$ is monotone decreasing, this implies the statement.
\end{proof}

Finally, we turn to a lower bound for the critical temperature.
We assume that the interaction potential is given by $\lambda V$,
with $V\leq 0$ not identically zero, and $\lambda>0$ a coupling
parameter.

\begin{proposition}\label{lowerboundoncrittemp}
  Let $V \in L^{3/2}$, with $V\leq 0$ not identically zero, and let 
  $\mu>0$. Then there exists a monotone increasing strictly positive
  function $g_{\mu, V}$, with $g_{\mu, V}(t) \sim e^{-1/t}$ as $t\to
  0$, such that the critical temperature for the potential
  $\lambda V$ satisfies the lower bound
\begin{equation}\label{lowerb}
T_c \geq g_{\mu, V}(\lambda)
\end{equation}
for $\lambda >0$.
\end{proposition}

\begin{proof}
  We have already shown in Thm.~\ref{potandcritical} that $T_c>0$ for positive
  $\lambda$. By Prop.~\ref{criticalbeta}, the fact that $V$ is
  non-positive and \eqref{uplowboundK} we have that
$$
1 /\lambda = \left\| |V|^{1/2}\frac 1{K_{\beta_c,\mu}}
  |V|^{1/2}\right\| \geq \left\| |V|^{1/2} \frac
  1{|p^2-\mu|+2/\beta_c} |V|^{1/2}\right\| \,.
$$
For any function
$\psi\in L^2$ with $\|\psi\|_2=1$ we therefore obtain
$$ 1/\lambda \geq \int \left|\widehat{\psi |V|^{1/2}}(p)\right|^2
\frac 1{ |p^2-\mu|+2/\beta_c}\, dp\,.
$$
The latter integral in monotone increasing in $\beta_c$ and
diverges logarithmically as $\beta_c \to \infty$ for appropriate
$\psi$ such that $\widehat{\psi |V|^{1/2}}(p)$ does not vanish
identically if $p^2 = \mu$. (See the proof of
Thm.~
\ref{potandcritical}(i) for an argument concerning the existence
of such a $\psi$.)  This implies the statement.
\end{proof}

\section*{Acknowledgments}
The authors would like to thank George Bruun, Thomas Hoffmann-Ostenhof
and Michael Loss for valuable suggestions and remarks.  C.H., E.H.,
and R.S. thank the Erwin Schr\"odinger Institute in Vienna for the
kind hospitality during part of this work. R.S. acknowledges partial
support by U.S. National Science Foundation grant PHY-0353181 and by
an Alfred P. Sloan Fellowship.

\appendix
\section{Motivation and Physical Background}\label{appa}

In this Appendix, we shall briefly explain the derivation of the BCS
functional defined in Definition~\ref{defF}. Our presentation follows
the one of Leggett in \cite{Leg}. Similar derivations can be found in
\cite{FW,MR}.  The BCS functional is obtained from several approximations and
assumptions on the underlying many-body problem. We assume the reader
to be familiar with the usual Fock space formalism in quantum
statistical mechanics.

Consider a system of spin $\half$ fermions confined to a cubic box
$\Omega\subset \R^3$, with periodic boundary conditions,
interacting via a two-body potential $V$. Assume, for simplicity,
that $V \in L^1(\R^3)$ so that it has a bounded Fourier transform, 
denoted by $\hat V(p)=(2\pi)^{-3/2}\int V(x) e^{-ipx}dx$. The
many-body Hamiltonian on Fock space is then given as
\begin{equation}\label{redham}
\Hr = \sum_{\sigma, k} k^2 \, a^\dagger_{k,\sigma} a_{k,\sigma} +
\frac{(2\pi)^{3/2}}{2|\Omega|} \sum_{k,p,q,\sigma,\nu} \hat V(k)
a^\dagger_{p-k,\sigma} a^\dagger_{q+k,\nu} a_{q,\nu}
a_{p,\sigma}\,.
\end{equation}
Here, $a^\dagger_{k,\sigma}$ and $a_{k,\sigma}$ denote the usual
creation and annihilation operators for a particle of momentum $k$ and
spin $\sigma$, satisfying the canonical anticommutation relations. The
momentum sums run over $k\in (2\pi/L) \Z^3$, where $L$ is the side
length of the cube $\Omega$, whereas the spin sums are over $\sigma\in
\{\uparrow,\downarrow\}$. Units are chosen such that $\hbar= 2m=1$,
$m$ denoting the particle mass.

Following \cite{BCS,Leg} we restrict $\Hr$ to BCS type states.
These are quasi-free states that do not have a fixed number of
particles. For details on the formalism, we refer the reader to
\cite{BLS}. Quasi-free states are linear maps $\rho$ from bounded
operators on the Fock space $\F$ to the complex numbers,
satisfying $\rho(A^\dagger A) \geq 0$, $\rho(I)=1$, and
$$ \rho(e_1e_2e_3e_4) = \rho(e_1e_2)\rho(e_3e_4) -
\rho(e_1e_3)\rho(e_2e_4) + \rho(e_1e_4)\rho(e_2e_3),$$ whenever
$e_i$ is either $a_{k,\sigma}$ or $a^\dagger_{k,\sigma}$.

Assuming both translation invariance and $SU(2)$ invariance for
rotations of the spins, the state $\rho$ is completely determined
by the quantities
$$\gamma(k) =\rho(a^\dagger_{k,\uparrow}
a_{k,\uparrow})= \rho(a^\dagger_{k,\downarrow}a_{k,\downarrow})\,,
\quad \varphi(k)
=\rho(a_{-k,\uparrow}a_{k,\downarrow})=-\rho(a_{-k,\downarrow}a_{k,\uparrow})\,.$$
All other expectation values of quadratic expressions are zero
under these symmetry assumptions. Note that $0\leq \gamma(p)\leq
1$ and $|\varphi(p)|^2\leq \gamma(p)(1-\gamma(-p))$ by Schwarz's
inequality and the canonical anticommutation relations. Since also
$\varphi(p)=\varphi(-p)$, this implies that $|\varphi(p)|^2\leq
\gamma(p)(1-\gamma(p))$.

The energy expectation value can be conveniently expressed in
terms of these quantities:
\begin{align}\nonumber\label{enfunct}
  \rho(\Hr) &=  2\sum_{k} k^2\,\gamma(k) + \frac {(2\pi)^{3/2}}{|\Omega|} \sum_{k,p}
  \hat V(k)\overline{\varphi}(k-p)\varphi(p)\\ &\quad + \frac{2(2\pi)^{3/2}}{|\Omega|}
  \hat V(0) \left(\sum_p \gamma(p)\right)^2 - \frac {(2\pi)^{3/2}}{|\Omega|}
  \sum_{k,p} \hat V(k)\gamma(p-k)\gamma(p)\,.
\end{align}
The first term in the second line is just a constant times the
square of the number of particles. The last term is the exchange
terms. Both these terms will be dropped in the sequel. This
truncations can be justified on physical grounds for interaction
potentials $V$ that are of short range such that $s$-wave
scattering is dominant \cite{Leg}.

The {\it von-Neumann entropy} of the quasi-free state $\rho$ under
consideration here, $S(\rho)=-{\rm Tr}_\F \rho\ln\rho$, can be
conveniently expressed in terms of the $2\times 2$ matrix
\begin{equation}
\Gamma(p) = \left(\begin{matrix}\gamma(p) & \overline\varphi(p)\\
\varphi(p) & 1- \gamma(p)
\end{matrix}\right).
\end{equation}
Note that $0\leq \Gamma(p)\leq 1$ as an operator on $\C^2$.  The
entropy is then given by (compare with \cite[Eq.~(2.c.10)]{BLS})
$$ S(\rho) = - 2 \sum_p {\rm Tr}_{\C^2}\left[\Gamma(p) \ln \Gamma(p)\right]\,.$$
The factor $2$ results from the $2$ spin components.

The {\it thermodynamic pressure}, $\Pp$, of a state at a fixed
temperature $T = \frac 1\beta \geq 0$ and chemical potential $\mu$
is defined as
$$\Pp(\rho) = \frac 1{|\Omega|}\left[T S(\rho) - \rho(\Hr -\mu\N) \right]\,.$$
Here, $\N = \sum_{\sigma, k} a^\dagger_{k,\sigma} a_{k,\sigma}$
denotes the number operator. Dropping the last two terms in
(\ref{enfunct}) and taking a formal infinite volume limit, one thus
obtains
\begin{align}\nonumber
\Pp(\gamma,\alpha):=\lim_{\Omega\to\R^3}\Pp(\rho) &= - \frac
2{(2\pi)^3} \int_{\R^3} \left( T\, {\rm
Tr}_{\C^2}\Gamma(p)\ln\Gamma(p) + (p^2-\mu) \gamma(p)\right) dp
\\ &\quad \nonumber -\frac 1{(2\pi)^{3}} \int_{\R^3} |\alpha(x)|^2
V(x) \, dx\,.
\end{align}
Here, $\alpha$ denotes the inverse Fourier transform of $\varphi$,
$\alpha(x) = (2\pi)^{-3/2} \int \varphi(p) e^{ipx} dp$. If we set
$\F_\beta= -\half (2\pi)^3 \Pp$ and replace $V$ by $2 V$ to slightly
simplify the notation, we are thus led to the definition of the BCS
model given in Definition~\ref{defF}.

\bigskip
Finally, we will briefly  explain in which sense $E(p)$ is an
effective dispersion relation and how it is related to an energy
gap between the ground state and the first excited state of the
system.

With $\varphi=\hat\alpha$ the Fourier transform of a maximizer of
the pressure functional at $T=0$, we introduce the
following quadratic Hamiltonian as an approximation to the full Hamiltonian $\Hr$:
\begin{equation}
{\mathbb{H}}_{\rm qu} =\sum_{\sigma, k} k^2\, 
a^\dagger_{k,\sigma} a_{k,\sigma} +  \frac{(2\pi)^{3/2}}{|\Omega|}
\sum_{ k,p} \hat V(k)\left( \overline\varphi(k-p)a_{-p,\uparrow}
a_{p,\downarrow}  + \varphi(p) a^\dagger_{p-k,\uparrow}
a^\dagger_{k-p,\downarrow}\right) .
\end{equation}
The Hamiltonian ${\mathbb{H}}_{\rm qu}-\mu \N$ can be diagonalized in
a standard way (see, e.g., \cite{MR}) via a Bogoliubov
transformation. Using the fact that $\varphi$ satisfies the gap
equation $E(k) \varphi(k)= - \frac 12 (2\pi)^{3/2} \sum_{k'} \hat
V(k-k') \varphi(k')$, where $E(k) = \sqrt{(k^2 - \mu)^2 +
  |\Delta(k)|^2}$ and $\Delta$ is determined from $\varphi$ via
(\ref{gal})--(\ref{bcsgapequationfiniteT}), this
leads to
\begin{equation}\label{reducedhamiltonian}
{\mathbb{H}}_{\rm qu}-\mu\N - \inf {\rm spec}\,\left(
  {\mathbb{H}}_{\rm qu} - \mu \N\right)=
\sum_{k} E(k) \left( c_k^\dagger c_k + d_k^\dagger d_k \right)\,.
\end{equation}
 The operators
$c_k$, $c_k^\dagger$, $d_k$ and $d_k^\dagger$ satisfy the
canonical anticommutation relations, and both $c_k$ and $d_k$
annihilate the BCS ground state. Equation
\eqref{reducedhamiltonian} thus explains why $E(k)$ is interpreted
as the quasi-particle dispersion relation and $\Xi =\inf_{k} E(k)$
as the {\it energy gap} of the system.


\end{document}